\documentclass[aps,pre,twocolumn,a4paper,floatfix]{revtex4}
\usepackage{epsfig}                                
\usepackage{amssymb}
\usepackage{amsmath}
\usepackage{graphicx}
\usepackage{color}

\begin{document}

\title{Extracting topological features from dynamical measures\\
 in networks of Kuramoto oscillators}

\bigskip
\author{Luce~Prignano and Albert~D\'{\i}az-Guilera}
\affiliation{Departament de F\'{\i}sica Fonamental, Universitat de Barcelona
08028 Barcelona, Spain}
\date{\today}

\begin{abstract}  
The Kuramoto model for an ensemble of coupled oscillators provides a paradigmatic example of non-equilibrium transitions between an incoherent and a synchronized state.
Here we analyze populations of almost identical oscillators in arbitrary interaction networks. 
Our aim is to extract topological features of the connectivity pattern from purely dynamical measures, based on the fact that in a heterogeneous network the global dynamics is not only affected by the distribution of the natural frequencies, but also by the location of the different values.
In order to perform a quantitative study we focused on a very simple frequency distribution considering that all the frequencies are equal but one, that of the pacemaker node. 
We then analyze the dynamical behavior of the system at the transition point and slightly above it, as well as very far from the critical point, when it is in a highly incoherent state. 
The gathered topological information ranges from local features, such as the single node connectivity, to the hierarchical structure of functional clusters, and even to the entire adjacency matrix.

\bigskip
\noindent
PACS number(s): 89.75.-k,  89.75.Hc, 05.45.Xt
\end{abstract}

\maketitle
 
\section{Introduction}
Nowadays, it is widely acknowledged that complex patterns of interaction are ubiquitous in nature as in society \cite{ba02}. 
Nonetheless, further research is required to completely understand 
how the topology affects the system dynamics \cite{n03a,blmch06}.
In particular how global dynamical properties are related with the units dynamics and the interactions between them. 
A unique answer cannot be provided since complex networks respond differently 
depending on the dynamical processes that take place on them \cite{indiana_book}.  

One of the most interesting of these macroscopically defined dynamical processes is synchronization, 
an emerging phenomenon in which populations of interacting units 
display a common periodic behavior \cite{prk01,okz07}. 
Indeed, understanding the role of connectivity in synchronization 
has been the subject of intense research in recent years \cite{adkmz08}. 
On the one hand, much work has focused on the generic properties of dynamical systems, 
mainly looking for necessary and sufficient conditions 
that would grant that a population of units under a set of simple dynamical rules 
is able to synchronize \cite{winfree80}. 
On the other hand, much progress has been made by studying precise models of phase oscillators, 
being one of the most paradigmatic the model proposed by Kuramoto \cite{kuramoto75, abprs05}, 
where the interaction between the units is proportional to the sine of the phase difference. 

In the present work, we will continue along this line and analyze a population of Kuramoto oscillators 
with a precise distribution of frequencies. 
The original work by Kuramoto and many subsequent studies considered that the oscillators, each coupled equally to all the others, had natural frequencies taken from a given distribution. The non-zero width of those distributions made the units follow different trajectories, whereas the interaction term made their phases approach. In fact and depending on the width of the frequency distribution, there is a critical value of the interaction strength above which the units tend to entrain their phases and hence leave the incoherent regime. If the natural frequencies of the oscillators are identical, a unique outcome is possible as the only attractor of the dynamics is a completely synchronized state in which all the oscillators end up in a common phase. And this occurs for any initial conditions and for any (connected) topology  \footnote{There are, however, some limitations to these results; for instance when the units are placed in regular lattices, such as one-dimensional rings, where other attractors different form the synchronized (equal phases) state, may arise \cite{da08}. }.

In systems with regular patterns of connectivity (including all-to-all)
the only complexity comes from the frequency distribution, 
whereas in more realistic (non-homogeneous) patterns, 
not only the frequency values matter but the precise location as well
\cite{bld09,blmch06}.

Here we will focus on a particular frequency distribution, 
one which is just one step away from the homogeneous case. 
Such distribution has identical frequencies for all oscillators except one. 
This singular oscillator, with a higher frequency than the rest, 
has received the name of \textit{pacemaker} and its effect in populations 
of Kuramoto oscillators has been analyzed\cite{km04, rm06}. 
In \cite{km04}, Kori and Mikhailov consider a special case 
where the pacemaker affects its neighbors but it is not affected by them; 
under these conditions they find numerically that the range of frequencies 
of the pacemaker for which the system can attain global synchronization 
depends on the "depth" of the network, defining the depth as the maximum 
distance from the pacemaker to peripheral nodes. 
Radicchi and Meyer-Ortmanns \cite{rm06} consider 
regular structures for which the conditions to synchronize can be 
analytically computed.

In this paper we use several properties of the heterogeneity induced by
the existence of the pacemaker to find useful relations between topology 
and dynamics. 
On one hand, by knowing the topology one should be able to infer the 
dynamical properties of the network. 
On the other hand, by measuring the dynamics some structural properties 
can be inferred, and this will be our purpose.

First, we use a similar procedure than the one used in \cite{km04} 
and \cite{rm06}, showing that there is a critical value for the 
frequency of the pacemaker above which the (frequency) synchronized 
state cannot exist. 
This is related with the existence of a synchronized solution 
(also exploited in \cite{mori04}) that applies to any subset of 
oscillators. We find, however, that from a practical point of view 
the most restrictive condition is usually for the equation of the 
pacemaker that involves its connectivity, and hence there is a clear 
relationship between the critical frequency and the pacemaker 
connectivity which can be used as an experimental measure of the degree.

In order to get more details on the network structure we analyze 
the system  above the critical value where correlations between 
dynamical evolution of the nodes appear. 
Such correlations enable to reveal the hierarchical organization 
and to recover the network connectivity.

The structure of the paper is as follows. 
First, in Sec.\,2, we characterize the coherent state and the transition to the incoherent one 
by means of a proper definition of the order parameter. 
Then, in Sec.\,3, we qualitatively analyze the behavior of the system 
when it is not in the frequency-locked state. 
Sec.\,4 is devoted to study the relation between local connectivity and the ability of the system 
to reach a synchronized (frequency-locked) state.
In Sec.\,5 we focus on the system slightly above the transition towards the incoherent state. 
We show that it is possible to perform some hierarchical analysis concerning the connectivity network. 
Finally, in Sec.\,6 we study the system far above the critical point, 
in a regime characterized by short range correlations
where it becomes easy to identify the nodes directly connected to the pacemaker. 
Thus the reconstruction of the whole connectivity pattern is accurate and fast.

\section{Synchronization and phase transition} 
In the original Kuramoto model \cite{kuramoto75,abprs05}, 
the phases of the oscillators evolve according to the following equation
\begin{equation}
\dot{\varphi}_i=\omega_i+\sigma\sum_{j=1}^N\sin(\varphi_j-\varphi_i),
\end{equation}
where $N$ is the total number of units of the system,
$\omega_i$ is the natural frequency of unit $i$, taken from a distribution, 
and $\sigma$ stands for the coupling strength. 
This case corresponds to a fully connected topology, i.e. each unit interacts with all the other ones. 
The ability of the system to reach a coherent state, for a given coupling strength, 
depends only on the width of the distribution of natural frequencies.

Here we want to consider arbitrary connectivity patterns. 
In this situation, the behavior of the system can no longer be understood in terms of 
the ratio between the distribution width and the coupling strength only. 
It is also relevant where the natural frequencies values are located, since on a generic interaction network 
nodes are not equivalent anymore. 

\begin{figure}[htbp]
\begin{center}
\includegraphics[width=\columnwidth]{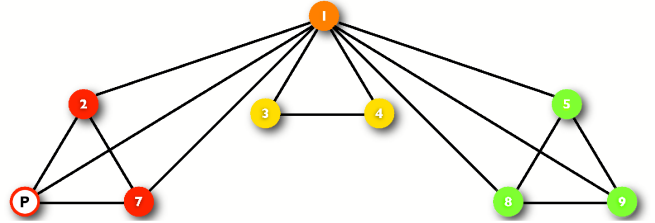}
\caption{(Color online) Hierarchic network that will be used as benchmark. 
In this particular setting the pacemaker is located on a peripheral node 
(marked as $P$) of degree $k_p\,$$=$$\,3$. 
The other nodes are grouped into sets using different colors. 
The elements of each set are topologically equivalent if we look at the network 
from the point of view of the pacemaker. 
Consequently their dynamical evolution is identical.}
\label{pdv}
\end{center}
\end{figure}

From now on we are using the 2 levels hierarchical network of 9 nodes represented in Fig.\,\ref{pdv}
as a benchmark and, when not otherwise stated, all the figures refer to that connection pattern.
This network has been presented in\,\cite{adp06b} as a very simple example 
of the class of deterministic scale-free hierarchical networks proposed by Ravasz and Barabasi in\,\cite{rb03}.
We choose this small regular connectivity pattern as a simple paradigmatic example showing general properties 
of the studied systems, since it makes easy to recognize the role of each node. 

%

Let us rewrite the equation for the evolution of the phases including a connectivity matrix $a_{ij}$ 
that is symmetric and takes values $1(0)$ if node $i$ and $j$ are connected (disconnected):
\begin{equation}
\dot{\varphi}_i=\omega_i+\sum_{j=1}^N a_{ij}\sin(\varphi_j-\varphi_i),
\label{dyn_eq}
\end{equation}
where we have rescaled time by setting $\sigma=1$.
Now we consider all the oscillators to have the same natural frequency (0 without loss of generality),
except one of them, called the pacemaker, whose frequency is $\omega\neq 0$. It is precisely this extremely simple choice of frequencies that enables to study the roles played by individual oscillators.

If a stationary state exists, then all the effective frequencies will take constant values 
and the following conditions have to be satisfied:
\begin{equation}
\sum_{j=1}^N a_{ij}\sin(\varphi_j-\varphi_i)=\Omega_i\ \ \ \forall i\neq p
\label{st_i}
\end{equation}
\begin{equation}
\omega+\sum_{j=1}^N a_{pj}\sin(\varphi_j-\varphi_p)=\Omega_p
\label{st_p}
\end{equation}
where $\{\Omega_i\}$ are the effective frequencies of the oscillators. 
Notice that summing up eqs. (\ref{st_i})-(\ref{st_p})  
the coupling terms cancel 
because of the symmetry of the interaction and it results in
\begin{equation}
\sum_{i=1}^N\Omega_i=\omega.
\label{claw}
\end{equation}

Looking at eqs.\,(\ref{st_i})-(\ref{st_p}) it is easy to recognize that 
there is an interplay between two effects. 
On the one hand the width of the frequencies distribution 
(in our present case this role is played by $\omega$ itself) 
tends to keep the evolution of the oscillators apart 
since each one follows its natural frequency. 
On the other hand, the interaction term makes them to approach their phases 
as well as their effective frequencies. 
Then we conclude that if the pacemaker natural frequency is 
small enough, 
the interaction term dominates and, after a transient time, 
all effective frequencies $\Omega_i$ will be identical
\begin{equation}
\Omega_i=\omega/N\ \ \forall i,
\label{omegas}
\end{equation}
including the pacemaker.
In this case we can say that the system is in a frequency-locked state, 
since all oscillators have the same frequency although the phases are not equal, 
because there is a coupling term (that of the pacemaker) that cannot vanish.

When increasing the pacemaker frequency $\omega$, 
some oscillators cannot keep the phase difference and the frequency-locked state is broken. 
The left-hand side of eq.\,(\ref{st_i}) is indeed bounded because of the sine terms, 
whereas the right term increases as the pacemaker frequency is increased. 
A similar conclusion can be  deduced from eq.\,(\ref{st_p}). 
Consequently, there will be a transition from a synchronized to an incoherent state. 
Thus we can define the critical value $\omega_p^c$ as the maximum value of the natural frequency of the pacemaker for which the system can attain global synchronization.

Such a transition for a population of phase oscillators is typically characterized 
by an order parameter $R$, defined through the equation:
$R\,e^{i\Psi}=\sum_je^{i\varphi_j},$ %
where $\Psi$ is a global phase (not constant) \cite{kuramoto84}.

\begin{figure}[htbp]
\begin{center}
\includegraphics[width=\columnwidth]{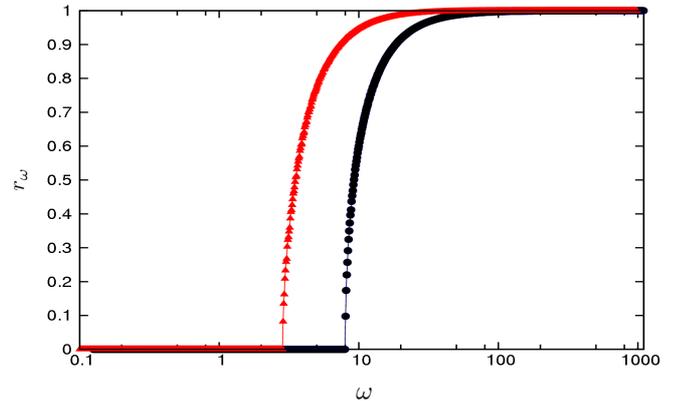}
\caption{(Color online) Order parameter (\ref{op}) as a function of the natural frequency of the pacemaker. 
Different curves correspond to different settings: 
($\bullet$) refers to the pacemaker located on node\,1 in Fig.\,\ref{pdv} ($k_p\,$$=$$\,8$),
($\blacktriangle$) to the pacemaker on node\,2 ($k_p\,$$=$$\,3$). 
The average value $\langle r_{\omega}\rangle_t$ for $\omega>\omega_p^c$ was calculated on a time window $\Delta t=100$.}
\label{OPmano}
\end{center}
\end{figure}

Anyway, in the present work, following \cite{bld09, nadal}, we adopt another order parameter 
that is a normalized measure of the effective frequency dispersion (standard deviation): 
\begin{equation}
r_{\omega}=\sqrt{\frac{\sum_{i=1}^N \left[\dot{\varphi_i}/\langle\omega\rangle-1\right]^2}{N-1}},
\label{op}
\end{equation}
where $\langle\omega\rangle$ is the average effective frequency of the oscillators population, 
a constant quantity always equal to $\omega/N$. 
According to its definition, $r_{\omega}$ takes values in the interval [0\,,\,1] (see Fig.\,\ref{OPmano}).
It should be noticed that, since above the critical frequency the system is not able to reach a steady state anymore, 
calculation of the order parameter (\ref{op}) requires to perform averages over an appropriate time window. 
Anyway, the value of $\langle\omega\rangle$ does not change because what we found in (\ref{claw}) is a general result,
even for instantaneous values of the effective frequencies.

To find the precise value of the critical frequency we apply the Newton-Raphson method (NR) and check, as a function of the frequency $\omega$, whether the synchronized solution of eqs. (\ref{st_i})-(\ref{st_p}) exists.
To simulate the dynamics of the system in the incoherent state ($\omega>\omega_p^c$)
we take as initial phases $\{\varphi_i(0)\}$ the stationary values of the differences provided by the NR solution for $\omega=\omega_p^c$. Eqs. (\ref{dyn_eq}) are numerically integrated
with Euler's Method (first-order), unless otherwise stated, at fixed time step $\delta t=10^{-2}$. 

\section{Incoherent state}
Above the critical frequency $\omega_p^c$ the system is no longer in a stationary state
and hence the effective frequencies are no longer constant.

\begin{figure}[htbp]
\begin{center}
\includegraphics[width=\columnwidth]{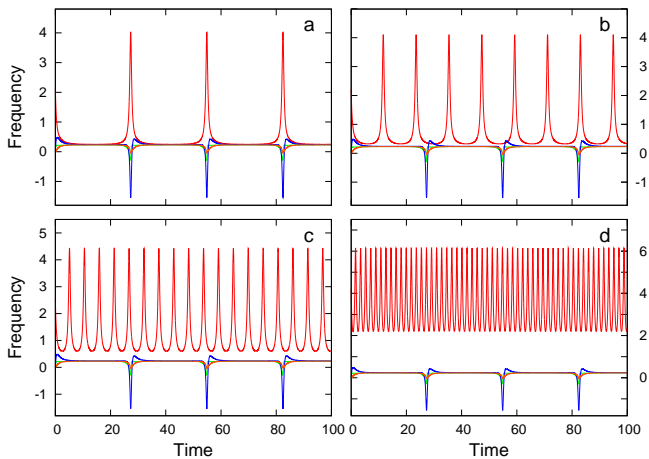}
\caption{(Color online) Effective frequencies above the critical point as functions of time. 0
That of the pacemaker (red, top curve), in this particular setting located on node $3$ in Fig.\,\ref{pdv}, 
is on average much larger than the others (lower curves). 
Panels a,\,b,\,c and d refer respectively to a pacemaker natural frequency value 
that is $1.01,\, 1.05,\, 1.2$ and $2$ times its critical value. 
Time starts after a transient lag $T_s=10$.}
\label{fvst}
\end{center}
\end{figure}

Numerical simulations show that, after a transient time, the system enters into a 
\textquotedblleft periodic\textquotedblright  state (see Fig\,\ref{fvst}).
The features of this periodic state are not affected by the initial conditions and they only depend on the pacemaker frequency and location.
It is precisely this fact that enables to infer topological properties from dynamical measurements.

\begin{figure}[htbp] 
\begin{center}
\includegraphics[width=8cm]{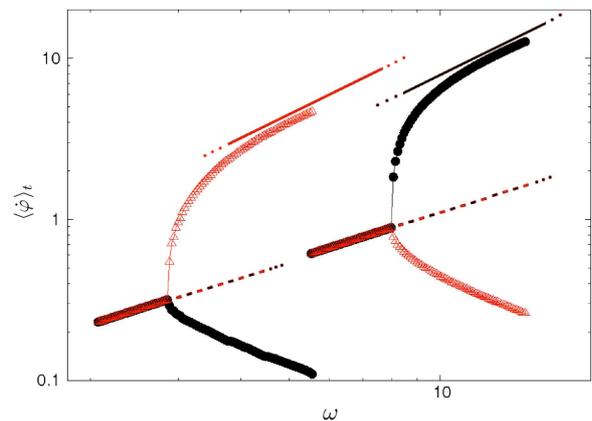}
\caption{(Color online) Average effective frequencies
as function of the natural frequency of the pacemaker $\omega$. 
The figure shows the behavior of two oscillators: node 1 (black circles) and node 2 (red triangles). 
On the right side the pacemaker is node 1, on the left one it is node 2. 
Initially the frequencies are synchronized and they increase linearly with slope $1/N$ (dashed line) 
as expected from eq.\,(\ref{omegas}). 
Then, when  $\omega$ reaches the critical value, 
that is different for different location of the pacemaker, they get apart. 
Far above the critical values, the average frequency of the pacemakers approach asymptotically 
a new reference line with slope 1 (solid line). 
The time averages were performed on a time window $\Delta t=100$.}
\label{forks}
\end{center}
\end{figure}

Fig.\,\ref{forks} summarizes what we have learned up to now, 
shedding light on some interesting details. 
The time average of the effective frequency of the pacemaker $\langle\dot{\varphi}_p\rangle_t$ 
and that of one of its neighbor $\langle\dot{\varphi}_j\rangle_t$ are plotted as functions 
of the pacemaker natural frequency. 
These quantities are calculated from numerical simulations 
taking into account appropriate time windows. 

Starting from small values of $\omega$, the picture shows 
how all the effective frequencies increase together linearly, 
following the reference line $\Omega_i=\omega/N$ defined by eq.\,(\ref{omegas}). 
Then, when $\omega$ reaches the critical value $\omega^c_p$, they do separate. 
Initially, the average effective frequency of the pacemaker goes through 
a more than linear increasing, while the others start decreasing, 
keeping their (average) values very close to each other. 
For even larger values, when $\omega\gg\omega_p^c$\,, 
Fig.\,\ref{forks} shows how the average effective frequency $\langle\dot{\varphi}_p\rangle_t$ 
tends to $\omega$, asymptotically  increasing along a new reference line 
with slope equal to $1$. 
At the same time, $\langle\dot{\varphi}_i\rangle_t$ for $i\neq p$ 
goes to zero, as required by the conservation law 
(\ref{claw}).

\section{Critical frequency and local topology}
In this section we explore the relation between the topology of the network 
and the value of the critical natural frequency of the pacemaker depending on the node where it is located. 

Let us begin by writing the equation for the pacemaker in the synchronized state. 
As a consequence of eq.\,(\ref{omegas}), we have
\begin{equation}
\omega+\sum_{j=1}^N a_{jp} \sin(\varphi_j-\varphi_p)=\omega/N.
\label{syncp}
\end{equation}
This equation links the natural frequency of the pacemaker 
to the constant values of the phase differences between it and its neighbors, 
when all the units are oscillating with the same effective frequency. 
Since the number of non-null terms $a_{jp}$ in the previous expression 
is given by the number of nodes connected with the pacemaker 
and $\sin(\varphi_j-\varphi_i)\in[-1,1]$, 
the degree (or connectivity) of the pacemaker is a bound for 
the absolute value of the sum in eq.\,(\ref{syncp}). 

Thus there is an upper bound for the critical frequency: 
\begin{equation}
\omega_p^c\leq k_p\frac{N}{N-1},
\label{fc}
\end{equation} 
where $k_p$ is the degree of the pacemaker. 
Indeed, any value larger than the right term in the inequality (\ref{fc}) 
is surely unable to satisfy eq.\,(\ref{syncp}) and hence the system is unable to be frequency synchronized.

Notice that we have obtained this bound by taking into account a single equation, that of the pacemaker. 
We can write for any oscillator the analogous of eq.\,(\ref{syncp}) as follows
\begin{equation}
\sum_{j=1}^N a_{ji} \sin(\varphi_j-\varphi_i)=\omega/N,\ \ \forall i\neq p.
\label{synci}
\end{equation}
It is easy to verify that no stricter condition can arise from any of these equations
\footnote{Applying the same argument to the eq. of the $i$-th node, we obtain $\omega_p^c\leq k_iN$, 
whose smallest possible value is $N$, that is the largest possible value for the bound (\ref{fc}).}. However, 
stronger bounds could exist due to the combination of eq.\,(\ref{syncp}) 
and some of  eqs.\,(\ref{synci}).

Let us consider a set of $(n+1)$ connected nodes, among which the pacemaker is included
\footnote{It is not necessary to take into account the groups that do not include the pacemaker, 
since the bound we obtain for $\omega_p^c$ 
summing up $n+1$ equations including the pacemaker
or the remaining $N-n+1$ (not including the pacemaker), is the same.} .
Labeling them by an increasing index $i=1,2,...,n+1=p$ and summing up their equations 
we obtain:
\begin{equation}
(n+1)\frac{\omega}{N} = \omega + \sum_{i=1}^{n+1}\sum_{j=1}^{N}a_{ij}\sin(\varphi_j-\varphi_i).
\label{n+1}
\end{equation}
If two nodes in the considered group are neighbors their respective interaction terms cancel each other. 
So the number of remaining terms of the sums in eq.\,(\ref{n+1}) is given by:
\begin{equation}
K_{out}=\sum_{i=1}^{n+1}k_i - \sum_{i,j=1}^{n+1}a_{ij}
\end{equation}
where $k_i$ is the degree of the $i-$th node and 
$K_{out}$ is equal to the number of links connecting the nodes of the considered set to external ones. 

Consequently, eq.\,(\ref{n+1}) can be rewritten as:
\begin{equation}
(n+1)\frac{\omega}{N} = \omega + \sum_{l=1}^{K_{out}}\sin(\phi_l),
\label{n+1bis}
\end{equation}
where $\phi_l=\varphi_j-\varphi_i$, 
being $i$ and $j$ connected nodes respectively inside and outside the group.

We are now able to write the expression of the upper bound 
for the critical frequency $\omega_p^c$ in a generalized form:
\begin{equation}
\omega_p^c \leq K_{out}\,\, \frac{N}{N-(n+1)} = N\,\, \frac{K_{out}}{N_{out}},
\label{new_up}
\end{equation}
where $N_{out}$ stands for the number of nodes not belonging to the considered set.
Eq.\,(\ref{new_up}) reduces to the previous upper bound if one chooses $n=0$.

In this way we can write a very large number of conditions, 
that is the number of the connected sets of nodes 
that include the pacemaker and which size ranges from $1$ to $N-1$. 
Among these, the strongest one is that for which the ratio $K_{out}/N_{out}$ takes its minimum value. 
This is a combinatorial problem, in principle very simple, but hard from a computational point of view, 
since the number of conditions grows at least exponentially with the network size.


Minimizing the ratio $K_{out}/N_{out}$ we find the strictest condition on $\omega_c^p$ 
that can be expressed in the form of a single equation. 
No other equation obtained as a linear combination of equations (\ref{st_i})-(\ref{st_p}) 
may provide a stronger bound. 
This condition is analogous to the necessary condition for global synchronization concerning the surface (here $K_{out}$) of any subset of nodes derived in \cite{mori04} for randomly distributed natural frequencies and generic oscillators.
However, these conditions are not sufficient. 
In our case,
it is not sure that the $K_{out}$ remaining sine terms of eq.\,(\ref{n+1bis}) 
are allowed to take their minimal values simultaneously. 
This kind of problems directly involves the sine functions arguments that may be not independent 
since they are differences between pairs of phases and we are dealing with a system of $N$ coupled equations. 
It may happen that two or more phases are tied among each other by a certain set of equations of the kind 
$f_i(\varphi_i, \{\varphi_{i_j}\})=0$ (where the nodes $\{i_j\}$ are neighbors of the node $i$). 
Consequently we cannot minimize the sum of sine terms on a hypercube $\left[0;\,2\pi\right]^{K_{out}}$ 
but we have to restrict ourselves on a hyper-surface of dimension $K_{out}-m$, 
where $m$ is the number of constrains. 
A system may experience this kind of difficulty (that we can regard as a kind of angles frustration) 
only if cycles are present and there is some anisotropy, and only when $1< k_p< N-1$.
Therefore, for a good number of regular connectivity patterns, as those analyzed in \cite{rm06}, 
there is not such a problem and
it is possible to analytically calculate the all set of values $\{\omega_c^{(p)}\},\ p=1,\dots,N$. 

As a simple analytically solvable network let us consider a Cayley tree 
with coordination number $z$, made up of $S$ shells. 
For each node it is indeed possible to single out a connected \textquotedblleft group\textquotedblright such that $K_{out}=1$, 
taking in all the nodes on the branch starting from the considered pacemaker. 
In this way we are minimizing the ratio $K_{out}/N_{out}$ 
so that we can consider the strictest equation among eqs.\,(\ref{new_up}). 
Moreover, since there is not any cycle, neither there are problems of angles frustration. 
Therefore, the obtained expressions give the correct values, not just bounds. 
In this way we obtain for the critical frequency:
$$\omega_c^{(s)}= N \frac {1}{N-\sum_{i=0}^{S-s}(z-1)^i},$$
where $s$ is the shell of the pacemaker.

\begin{figure}[t]
\begin{center}
\includegraphics[width=\columnwidth]{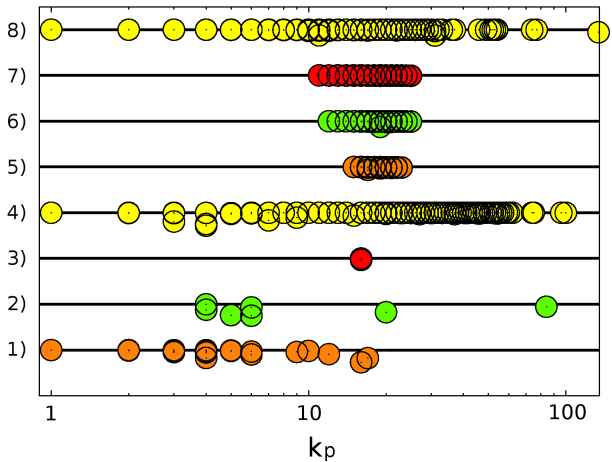}
\caption{(Color online) Critical frequency of a pacemaker as a function of its degree, for a set of networks. We have divided the critical frequency by the degree and by $N/(N-1)$ such that the bound given by eq.\,(\ref{fc}) is 1. We have shifted the data for the different networks and the horizontal lines are the reference (equal to 1) for each case. From bottom to top the networks are: 1) Zachary club social network \cite{gn02} used in community detection applications;
2) hierarchical network of 125 nodes and 3 levels \cite{rb03};
3) network of 4 communities of 32 nodes each used as benchmark in community detection algorithms \cite{gn02} where all the nodes have the same degree;
4) network of jazz bands \cite{gz06}; 
5-6-7) three networks of three levels of community structure used to relate topological and temporal scales in synchronization \cite{adp06a};
8) $Caenorhabditis$ $elegans$ neural network \cite{afg09}.}
\label{pacemakers}
\end{center}
\end{figure}

Even though in real complex networks it is not so easy to calculate the $\{\omega_c^p\}$, 
we have empirically verified that only in few cases the critical frequency 
is much smaller than its first upper bound (\ref{fc}).
This can be clearly observed in Fig.\,\ref{pacemakers}, 
where we plotted the ratios between the real critical values and the corresponding upper bound, 
for every choice of the pacemaker in several networks.

The accuracy of this estimation enables us  to use it in the opposite direction, i.e. to get an estimation of the pacemaker degree from an experimental measure of the critical frequency.
We can invert eq.\,(\ref{fc}) obtaining:
\begin{equation}
k_p\geq \omega_p^c\frac{N-1}{N},
\label{kc}
\end{equation} 
but, since the right term is not an integer, the smallest allowed value for $k_p$ is  
\begin{equation}
k_p^*=\left[\omega_p^c\frac{N-1}{N}+1\right], 
\label{kce}
\end{equation} 
where $[x]$ stands for the integer part of $x$. 
We can conclude that eq.\,(\ref{kce}) gives the correct value of $k_p$ whenever
$$\omega_p^c\in \left[(k_p-1)\frac{N}{N-1},k_p\frac{N}{N-1}\right].$$
This fact implies that the estimator (\ref{kce}) for the degree of the pacemaker is very reliable.
Indeed, it only fails 
when the critical frequency is really smaller than its bound (\ref{fc}).

\section{Slightly above the critical point}
In this and in the next section we translate 
the rich dynamical information that the system provides in the incoherent state 
into useful topological information.
Here we focus on the behavior of the system slightly above the critical point, 
while in Sec.\,6 we will analyze the system when the natural frequency of the pacemaker 
is many time larger than its critical value. 

We are interested in estimating how much similar two nodes are 
from a global topological perspective. 
For this purpose we need to define an appropriate correlation function, 
able to relate the dynamical responses of pairs of oscillators. 

Looking for the expression of a good correlation function, we 
get no help from the average values $\langle\dot{\varphi}_i\rangle_t=\int_0^{\infty}\varphi_i(t)dt$. 
Indeed, in this regime, all the oscillators, except the pacemaker, 
have the same average effective frequency.
On the contrary, it can be useful to look at the difference between instantaneous values. 
We measure the frequency of every oscillator at each time, inside a suitable interval. 
In order to define a correlation, that is a quantity that has to be non-negative 
and symmetric with respect to the two nodes indexes $i$ and $j$, 
it is reasonable to start from a power of the absolute value of the difference 
$|\dot{\varphi}_i^{(p)}(t)-\dot{\varphi}_j^{(p)}(t)|$, where $(p)$ stand for
the pacemaker that induces the considered dynamical evolution.  
Therefore, we propose
$$c^p_{ij}(t)= 1-\sqrt{\frac{|\dot{\varphi}_i^{(p)}(t)-\dot{\varphi}_j^{(p)}(t)|}{\omega}}.$$
Dividing by $\omega$ makes that the argument of the root is less than $1$ because, 
even if the frequencies may take negative values (see Fig.\,\ref{fvst}), 
the condition $|\dot{\varphi}_i(t)|\ll\omega$ always holds. 

The period of the effective frequencies oscillation depends on which node is the pacemaker. 
Then, in order to compute averages on time that are really independent from the considered interval, 
we have to choose a time window many times larger than the oscillation period. 
Furthermore, since $|\dot{\varphi}_i-\dot{\varphi}_p|\gg |\dot{\varphi}_i-\dot{\varphi}_j|$  for any $i,j \neq p$, 
we decide to exclude these contributions, 
taking into account only terms of the kind $c^p_{i,j}$ where $i\neq p$ and $j \neq p$. 
 
Finally, in order to remove the dependence from the index $p$ 
we have to average on all the possible pacemakers.
Summarizing in a compact expression, our correlation function can be written as follow:
\begin{equation}
c_{ij}=1-\frac{1}{N-2}\sum_{p=1\ p\neq i,j}^N\frac{1}{t_1-t_0}\int_{t_0}^{t_1}\sqrt{\frac{|\dot{\varphi}_i^{(p)}(t)-\dot{\varphi}_j^{(p)}(t)|}{\omega}}dt
\label{corr}
\end{equation}

\subsection{Hierarchical organization}
\begin{figure}[htbp] 
\begin{center}
\includegraphics[width=\columnwidth]{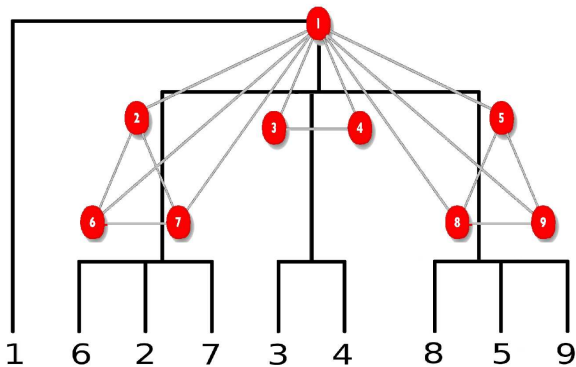}
\caption{(Color online) The network of Fig.\,\ref{pdv} and its 
corresponding dendrogram. 
Correlations are calculated averaging on a time window $\Delta t=60$, after a transient lag $T_s=10$, for $\omega=1.1\cdot\omega_p^c$.}
\label{dendro9}
\end{center}
\end{figure}

Once we have obtained the correlation matrix 
we can proceed to some hierarchical analysis.
In the present work we use the standard 
Unweighted Pair Group Method Average (UPGMA)\,\cite{snso73} 
algorithm to compute such diagrams.
What we find out is a hierarchy of dynamical communities, 
whose meaning is immediately understandable in the case of small
networks, such as our benchmark in Fig.\,\ref{pdv} (see Fig.\,\ref{dendro9}). 

Obviously, this simple network does not need 
any analysis to obtain its hierarchical organization, 
but this methodology can be very useful 
when applied to functional hierarchical network.

As a paradigmatic example, 
let us consider the corticocortical network of the cat at the macroscopic level. 
We look at each cortical area as a basic unit, modeling it as a Kuramoto oscillator,
finding out similar results as in\,\cite{zzk06,zzzhk06}.

\begin{figure}[htbp] 
\begin{center}
\includegraphics[width=\columnwidth]{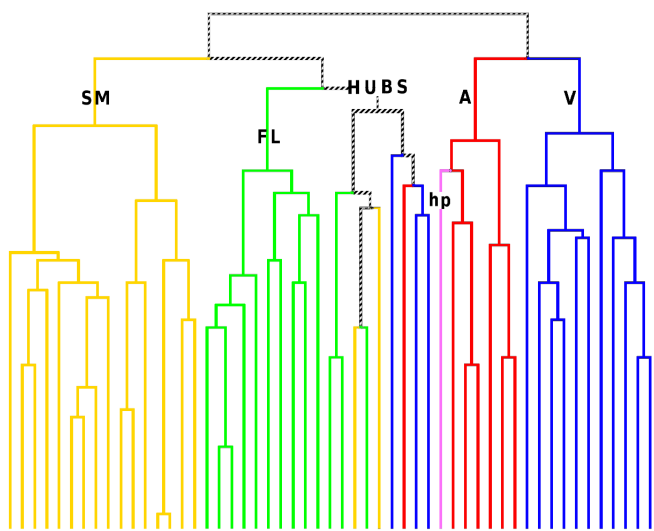}
\caption{(Color online) Dendrogram of the cortical brain network of the cat. Different colors correspond to
different sub-systems: the fronto-limbic (FL), the somatosensory-motor (SM), 
the auditory (A) and the visual (V).
The rich-club is labeled with “HUBS”, while the branch indicated with the label “hp” (pink) is the area that belongs to the hippocampus and it is out of its place. Correlations are calculated averaging on a time window $\Delta t=100$, after a transient lag $T_s=10$, for $\omega=1.1\cdot\omega_p^c$.}
\label{catree}
\end{center}
\end{figure}

In Fig.\,\ref{catree} we show that, 
going down along our dendogram starting from the root, 
it is possible to recognize two communities clearly separated. 
Then, the right branch splits up into two parts and the left one undergoes 
into two subsequent bifurcations, so that it is possible 
to identify three groups of nodes on it. 
At this level we have five communities. 
Four of them correspond to well known physiological sub-systems:  
the fronto-limbic (FL), the somatosensory-motor (SM), 
the auditory (A) and the visual (V).
The fifth one (HUBS) is composed - except for a single area \footnote{The cortical area that is not a super-hub 
is a border area that can be seen as a hub 
only joined with another one very similar to it, 
but anyway regarded as a super-hub in itself.}- 
by super-hubs, 
sometimes considered as a meta-community (rich-club) \cite{zzk06,zzzhk06}. 
The most relevant aspect of our hierarchical analysis is that 
there is no way to recognize this meta-community 
if the dendrogram is constructed by means of static methods.
Neither it can be obtained throughout  correlation matrices 
generated from the adjacency matrix using, 
for instance, Pearson's Coefficient\,\cite{rn88}. 
Nor these nodes emerge as a community when the modularity function is maximized.
Indeed, maximizing the modularity we obtain as an optimal partition 
the same 4 groups corresponding to the 4 physiological sub-systems.



In general, complex networks can be organized, and thus analyzed, at different hierarchical levels. For social networks it is very important that a group is tight, so that the multiple connections within the group give rise to the concept of community. On the contrary, in biological networks the most crucial concept is function rather than connectivity \emph{per se}. Therefore, methods that rely on the connections within groups and maximize modularity will not be enough to identify biological units, based primarily on function \cite{sus07,jdmmh11}. In this case, our method, which analyzes the dynamical correlation between units, provides a better approach to infer functional relationships.

One of the known problems of the methods commonly used for detecting community structures in complex
networks is the existence of the so called resolution limit, found by Fortunato and Barthelemy \cite{fb07}. This
issue is related to the impossibility for the methods based on modularity optimization to go beyond
certain resolution which is related to the community size and to the number of links between
communities. The paradigmatic example of the problem is a network formed by "cliques" (small groups
of totally connected nodes) which are very sparsely connected among them. We have checked such
structures and found that dynamically the correlations are very strong within the cliques and not
among nodes belonging to different modules, showing that our method detecting the hierarchical
organization is not affected by the resolution limit problem.

\subsection{Recovering network topology}
Let us now take a step backward and recover something we had previously discarded. 
In the sum of equation (\ref{corr}) we had excluded terms in which one of the indexes was equal to $p$ 
since they were \textit{heterogeneous}. So $c_{ij}=\frac{1}{N-2}\sum_{p\neq i,j}c_{ij}^p$. 
Anyway, also the set of elements $c_{pj}^p$, $p=1,\dots,N $ contains information. 
We may ask ourselves which are the oscillators most strongly correlated with the pacemaker and if they share some topological  property. 
The simplest hypothesis is that the set of $k_p$ largest $c_{pj}^p$ identifies the neighbors of the pacemaker. 
This is reasonable since, even if the pacemaker is very weakly correlated with the rest of the oscillators, 
coefficients $c_{pj}^p$ are not uniform and the topological distance is the most immediate quantity 
we may suppose this variability is related to. 
In Sec\,4 we showed how to find out an estimator of the degree of each node from the critical frequencies. 
Thus if we are able to select the possible neighbors we would be in principle able to reconstruct the entire network. 

The first problem we face in the attempt to validate this hypothesis is that 
our list of likely neighbors gives us an asymmetric and weighted adjacency matrix, 
which elements are 
\begin{eqnarray*}
a'_{pj_i}&=&c_{pj_i}^p\ \ \textrm{for}\ i=1,\dots,k_p^*,\\
a'_{pj_i}&=&0\ \ \textrm{for}\ i=k_p^*+1,\dots,N,
\end{eqnarray*}
where $k_p^*$ is the estimator for the degree of the pacemaker given by\,(\ref{kce}) and
$c_{pj_i}^p>c_{pj_l}^p$ whenever $i\leq k_p^*$ and $l>k_p^*$.

Moreover $a'_{mn}\neq a'_{nm}$ since generally speaking $c_{mn}^m \neq c_{mn}^n$.
Therefore we have to remove the weights and to symmetrize this matrix. 
Here we propose an algorithm to perform this task that is at the same time simple and efficient. 
It consists in four steps. 

1) Symmetrize the matrix in the usual way: $a^s_{mn}=(a'_{mn}+a'_{nm})/2$; 

2) Compute a list of temporary degree $k'_n\geq k_n^*$ as the number of non-null elements $a^s_{nm}$;

3) Order all the non-zero values $a^s_{nm}$ in a list, from the smaller to the larger; 

4) Check which ones among the corresponding likely links have to be removed,
starting from the weakest one. 

We proceed as follows:
given a pair of nodes $m$ and $n$ whose link is the weakest one, 
if and only if $k'_m>k_m^*$ and $k'_n>k_n^*$ we remove that link, 
setting $a^s_{mn}=a^s_{nm}=0$. 
In this case both $k'_m$ and $k'_n$ are reduce by one unit. 
Otherwise we go to the next link, 
going on along the entire list, till the strongest link.

This method roots in the hypothesis, empirically very well verified, 
that the matrix $a_{mn}^s$ contains all the links of the real network, 
plus a number of \textit{false positive} ones, 
i.e. that there is no \textit{false negative} link. 
Thus we need just to remove, never to add edges. 

Moreover it works properly only if our estimators $\{k_n^*\}$ of the actual degrees $\{k_n\}$ are correct, 
otherwise we may make additional errors. 
Fortunately it is a very infrequent problem. 
The sole hypothesis we make is that the probability for a link 
of being a \textquotedblleft false\textquotedblright one 
is a monotonously decreasing function of the correlation between the nodes it joins.

Finally, the method does not ensure that in the final estimated network 
$k'_n=k_n^*$ $\forall n$ 
because it is possible that even if $k'_n>k_n^*$, 
the $n$-th oscillator has no possible neighbor 
which temporary degree is larger than its estimated one. 
Sometime this fact may cause new errors,
some others it acts as a compensation of the underestimation 
of the real degrees.

In order to quantify how good a reconstruction is, we introduce the following error definition:
$$err\%=\frac{Fp+Fn}{L}\cdot100,$$
where $Fp$ and $Fn$ are respectively the number of false positive (spurious) and false negative (missing)
links in the reconstructed network, and $L$ is the number of edges in the original connectivity pattern.
Globally speaking, we can state that our method allows for a reconstruction 
of an arbitrary connectivity pattern with a good precision.
Taking into account the networks in Table\,\ref{tab}, on average we have $err\%=6.5$.

Among these networks there are artificial as well as real connectivity patterns.
They were selected to be representative of several classes of networks, 
including hierarchical as well as not hierarchical, 
with and without community structure, regular and not regular.
For this reason, the average error calculated on this set of benchmarks can be considered as a good estimator of the
accuracy of the proposed reconstruction method when applied on a given unknown connectivity pattern.
\begin{table}[htbp]
\begin{center}
\begin{tabular}{|c|c|c|c|c|c|c|c|}
\hline 
N & L & K$_{err}$ & L' & Fp/Fn & L'$_r$ & Fp/Fn & err\%\tabularnewline
\hline
\hline 
9 & 15 & 0 & 15 & 0/0 & 15 & 0/0 & 0\tabularnewline
\hline 
18 & 24 & 0 & 24 & 0/0 & 24 & 0/0 & 0\tabularnewline
\hline 
25 & 66 & 0 & 82 & 16/0 & 66 & 0/0 & 0\tabularnewline
\hline 
34 & 78 & 7 & 99 & 27/6 & 75 & 7/10 & 21.8\tabularnewline
\hline 
48 & 64 & 0 & 64 & 0/0 & 64 & 0/0 & 0\tabularnewline
\hline 
53 & 391 & 0 & 445 & 53/0 & 392 & 5/4 & 2.3\tabularnewline
\hline 
125 & 394 & 33 & 475 & 81/0 & 383 & 1/12 & 3.3\tabularnewline
\hline 
128 & 1024 & 0 & 1060 & 57/21 & 1026 & 36/34 & 6.8\tabularnewline
\hline
256 & 2311 & 0 & 3223 & 1040/128 & 2324 & 259/246 & 21.8\tabularnewline
\hline
256 & 2301 & 0 & 2851 & 607/57 & 2312 & 116/105 & 9.6\tabularnewline
\hline
\end{tabular}
\caption{Results of the reconstruction on several networks. 
On the columns we list: the size of the system (N), the total number of links in the original network (L), the total error in the estimation of the degrees (K$_{err}=\sum_{i=1}^N|k_i-k_i^*|$), the total number of links in the reconstructed network before the removal of exceeding links (L'), the number of false positive (Fp) and false negative (Fn) links in this network, the same for the final reduced network (L'$_r$, Fp/Fn) and the final total error (err$\%$). 
From the first row, the networks are: our usual benchmark \cite{adp06b}, ring of 6 cliques of 3 nodes \cite{fb07}, hierarchical network of 25 nodes and 2 levels \cite{rb03}, Zachary club social network \cite{gn02}, ring of 16 cliques of 3 nodes \cite{fb07}, cortical brain network of the cat \cite{sbhoy99}, hierarchical network of 125 nodes and 3 levels \cite{rb03}, network of 4 communities of 32 nodes each \cite{gn02}, 2 networks of 3 levels of community structure \cite{adp06a}.}
\label{tab}
\end{center}
\end{table}

\section{Far from the critical point}
\begin{figure}[htbp]
\begin{center}
\includegraphics[width=\columnwidth]{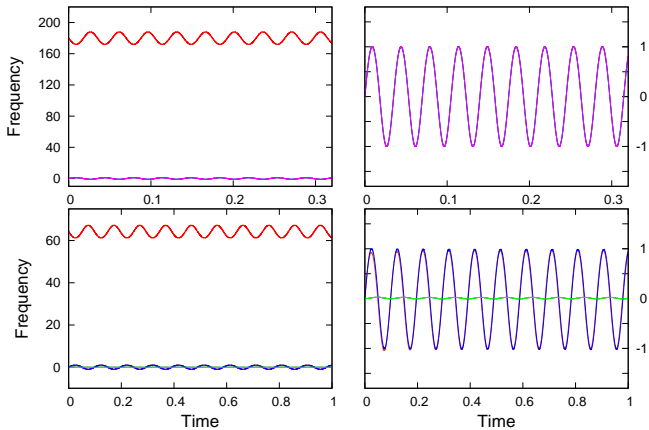}
\caption{(Color online) Effective frequencies as function of time far above the critical point ($\omega=20\cdot \omega_p^c$). Plots refer to the same network used for the previous pictures, in the case of 2 different choice of the pacemaker: node 1 ($k_1=8$) above; node 2 ($k_2=3$) below. On the left hand side we plotted the frequency of all the nodes in the network. On the right one the scale has been changed and the pacemakers are left out. Notice how above, where all the nodes are neighbors of the pacemaker, we may observe a unique curve. On the contrary, below there are 2 different kinds of oscillations. The largest ones are those of the neighbors of the pacemaker, the others belong to the oscillator not directly connected to it. Time starts after a transient lag $T_s=1$. The integration time step used is $\delta t=10^{-4}$.}
\label{far}
\end{center}
\end{figure}

Far above the critical point the system behaves quite differently.  
As clearly shown in Fig.\,\ref{far} all the units are characterized by effective frequencies that, 
after a transient time, oscillate around precise values that  
are equal to their own natural frequency, as can be seen in the left panels of Fig.\,\ref{far}. 
From this point of view, by increasing the natural frequency of the pacemaker the coupling 
is less and less important. 
But, in any case, there are still reminiscences of the interactions  
since the amplitudes of the oscillations decay very fast with the distance from the pacemaker. 
Indeed, the frequencies of the neighbors of the pacemaker oscillate with an amplitude 
that is roughly $A_{neigh}\simeq2$, while all the other oscillators are almost at rest if compared with them. 
These conditions allow us to recognize the neighbors of a given pacemaker even if we do not know how many they are.
Therefore, we may define a simplified correlation function that better suits this situation and that only connects each pacemaker with its neighbors:
\begin{equation}
c^{F}_{pi}=\frac{\max_t(\dot{\varphi}_i(t))-\min_t(\dot{\varphi}_i(t))}{\max_t(\dot{\varphi}_p(t))-\min_t(\dot{\varphi}_p(t))}=\frac{A_i}{A_p}.
\label{cfa}
\end{equation}
Previous expression is the ratio between two positive terms (amplitudes) and it is equal to $1$ for $i=p$.

On any connectivity pattern, the values $c_{pi}^{F}$ are distributed along a stair 
whose highest step is easy to identify even if we consider short time windows. 
The transient time, indeed, is always very short in this regime.
We do not need any more to completely reconstruct the entire connection topology.

All we have to do is 
to compute the values $c^{F}_{pi}$ for each pacemaker. 
After finding out the maximum values $\max_{i \neq p}c^{F}_{pi}$ $\forall p$, 
we choose an appropriate threshold, say $0.5$. A node $j$ will be a neighbor of the pacemaker $p$ 
if $c^{F}_{pj}/(\max_{i \neq p}c^{F}_{pi})\geq 0.5$. 
Now we are able to construct a connectivity matrix. 

Let us notice that in this case there is no need for symmetrization since the adjacency matrix constructed in this way 
is already symmetric because this method is based on a reliable general property 
that holds for any connectivity pattern.
\begin{figure}[htbp]
\begin{center}
\includegraphics[width=\columnwidth]{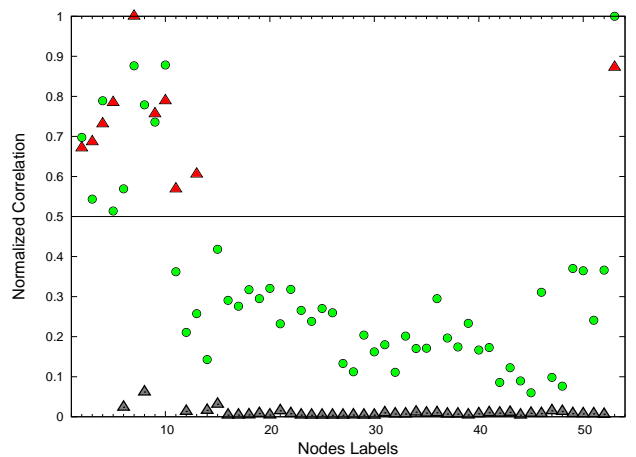}
\caption{(Color online) Normalized correlations of the cortical brain network of the cat for pacemaker on node 1 ($k_p=10$). The ($\bullet$) are the correlation values calculated through eq.\,(\ref{corr}) for $\omega=1.2\cdot\omega_p^c$ ($\Delta t=100$, $T_s=20$). The ($\blacktriangle$) correspond to the correlations given by expression\,(\ref{cfa}) when $\omega=20\cdot\omega_p^c$, calculated on a time window $\Delta t=1$ and waiting a transient time $T_s=0.1$. All the values have been divided by the maximum on each set (excluding the auto-correlation). Notice that while in the first case there is an almost continuous spectrum of values, in the second one it is easy to identify a group of points (in red) above the line at $0.5$ clearly separated from the rest. Those are the 10 neighbors of node 1.}
\label{cfacat}
\end{center}
\end{figure}
The use of a threshold is therefore in principle unnecessary, since all the neighbors 
have the same amplitude of the frequency oscillation, when the pacemaker natural frequency is above a certain value. 
But, since this value is not know \textit{a priori} and it may be very large if the distribution of the degrees 
among the neighbors of the pacemaker is very wide, it is useful from an empirical point of view. 
It is important to stress that, even if we are still in a regime where some degree of $heterogeneity$ among the neighbors 
is conserved, there is no chance to make any error in the recovered topology. Indeed, the amplitudes 
of the frequency oscillations of oscillators not directly connected to the pacemaker are at least one order of 
magnitude smaller than those of its neighbors (see Figs.\,\ref{far} and \ref{cfacat}). 
By means of this method all the topologies considered in Table\,\ref{tab} are properly  reconstructed, without errors.

In addition, not all nodes need to be considered as pacemakers.
While the method discussed in Sec. V-B 
requires  to perform dynamical measures for every possible location of the pacemaker, 
for the current description this is not necessary.
Indeed, we can look for the neighbors of a number $N'<N$ of pacemakers in order to get all the connections in the considered network. 
From an experimental point of view, adopting the conceptual framework proposed in\,\cite{timme07}, 
we may consider the choice of a certain pacemaker as the application of a drift on a given unit in a system of identical coupled oscillators.
This means that it is possible to solve the problem with less than $N$ experiments.

\begin{figure}[ht]
\begin{center}
\includegraphics[width=\columnwidth]{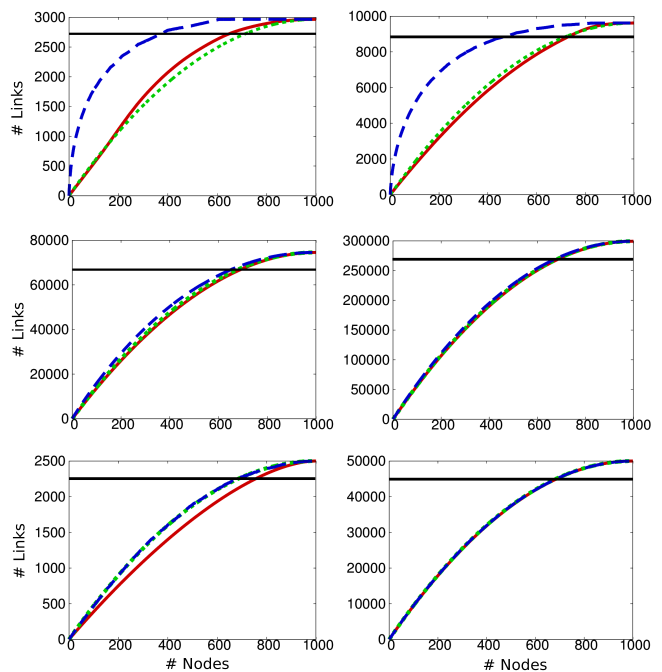}
\caption{(Color online) Average number of reconstructed links as a function of the number of nodes we considered as pacemakers (number of trials). 
From the top to the bottom, the considered networks are: a pair of Barabasi-Albert networks, respectively with parameter $k\,$$=$$\,3$ (left) and $k\,$$=$$\,10$ (right); a pair of Erdos-Reyni graphs with average degree equal to $15$ (left) and $60$ (right); a pair of random regular graphs with degree $5$ (left) and $100$ (right). The size is $N=1000$ for all of them.
Different lines corresponds to different selection algorithms. Blue dashed lines stand for the ordered sequence on the basis of the critical frequencies values; the red solid ones for the random walk; the green dots ones for random extractions. Both the random walk and the random extractions are averaged over 1000 samples. The horizontal black line marks the $\%90$ of links: notice how in any case we never need more than $\%70$ of the nodes in order to reconstruct $\%90$ of the links, decreasing to a $\%30-\%40$ in the case of the ordered sequence for scale-free networks.
Correlations are computed under the same conditions as those of Fig.\,\ref{cfacat}.}
\label{NL}
\end{center}
\end{figure}

The criterion for choosing the ordered sequence of nodes on which we  locate the pacemaker can vary. We may operate a random extractions, or we may start from a randomly chosen node and then move to one of its neighbors along a random walk.
Another option, much more convenient especially in the case of scale-free networks, can be adopted if the critical frequencies associated
to each oscillator are known. We can ordered the nodes according to decreasing critical frequency, starting from the highest one.
In this way we proceed from larger to smaller (estimated) degrees, taking an important advantage if the degrees distribution is not uniform and
there are hubs. The hubs, indeed, provide information about a large number of links by means of very few experiments (Fig.\ref{NL}).\\

\section{Conclusions}

Systems of non-identical Kuramoto oscillators have been recently shown to display a degree of synchronization that depends strongly on the topology of the underlying complex network. Here, these dynamical properties, in particular by setting different types of correlations between the dynamical evolution of the oscillators, have been used to gather information on the connectivity patterns. Remarkably, this is the case for most experimental situations, where the a priori unknown connectivity of a particular network is inferred from purely dynamical measurements.


When the oscillators are identical (all of them having the same natural frequency) any topological configuration has a unique attractor, which is the complete synchronized state; synchronized meaning that the oscillators end up in such a state that all effective frequencies and phases are identical.
This state does not offer any information about the topology.
We perturb this setting by allowing one of the oscillators to have a different natural frequency than the rest. This unit is called the pacemaker of the network. Such perturbation causes that the final state is no longer phase-synchronized. 
But if the natural frequency of the pacemaker is not very different from the value of the rest of the population, the system still will keep a certain degree of synchronization, since the whole system can evolve with the same  effective frequency.
However, if the frequency difference becomes larger, the system will be unable to find any kind of synchronization.
The threshold between the former case and this latter is a well defined value, which is strictly dependent on the location of the pacemaker in the network.  In this context, we can use the correlations between the effective frequencies of the  oscillators in such incoherent state
to reproduce the network connectivity. 

Moreover, we show that the dynamical correlations in different situations, whether close of far from the critical point, provide complementary information on the network:

\begin{enumerate}
\item Working around the critical point we are able to estimate the degree of each pacemaker merely by its critical frequency.
\item Slightly above the transition point
the hierarchical structure of the whole network (related to functional modules)  
can be obtained from the correlations between effective frequencies. 
A further refinement enables to recover the whole connection network with a good degree of accuracy.

\item
Far above the critical point
it is possible to recognize which are the oscillators that are directly connected to an individual pacemaker from a very short measurement of the time evolution of the effective frequencies. 
In this way we can recover the connectivity pattern and 
this method turns out to be much more precise and more efficient than the previous one. 

\end{enumerate}

In summary, this paper deals with different approaches relating dynamical properties of individual nodes to the topology of the network. The topological properties inferred from dynamics can be local (the existence of a link between two nodes) as well as global (hierarchical organization of the nodes in the functional network). 
In particular, for a scale-free network and if the node degrees are known (or have been estimated from the critical frequencies),  considering 30\% of the possible pacemakers, always selecting the most connected nodes, will be enough to reconstruct approximately 90\% of the links.

Other papers have considered the reconstruction of the network 
from dynamical information. 
Similar to our proposal with specific targets, Tegner et. al. 
\cite{tyhc03} analyzed the dynamical response of a gene-regulatory 
network by changing expression levels of particular genes. 
On the contrary, Di Bernardo et. al. \cite{db-c05} considered the 
global effect of different types of perturbations to infer the network 
topology. This approach has been followed recently also by Gorur Shandilya 
and Timme in \cite{gst11}, where it is assumed that there is some information 
about the dynamical evolution of the isolated units and about the coupling. 
Our method, based on the change of the frequency of a single unit 
and how it enhances correlations among the nodes, can be more effective 
in oscillatory systems. 
In any case, for practical purposes the method chosen will depend on 
the specific details of the experimental setup and even a combination 
of different ones can be the most appropriate.

\acknowledgments
The authors thank G. Zamora-Lopez for helpful discussions.
L.P. is supported by the Generalitat de Catalunya through the FI Program.
While part of this work was performed, A.D-G. was supported by Ministerio de Educación y Ciencia (PR2008-0114).
This work has been supported by the Spanish DGICyT Grants FIS-2006-13321 and FIS2009-13730, and by the Generalitat de Catalunya 2009SGR00838.

\bibliographystyle{apsrev}	
\bibliography{myrefs}		
\end{document}